%% file: main.tex
\documentclass[runningheads]{llncs}
\usepackage[T1]{fontenc}
\usepackage{graphicx}

\usepackage{hyperref}
\usepackage{amsmath}
\usepackage{amssymb}
\usepackage{xcolor}
\usepackage[english]{babel}
\usepackage{tcolorbox}
\usepackage{orcidlink}
\usepackage{adjustbox}
\usepackage{booktabs}
\usepackage{multirow}
\usepackage{colortbl}

\usepackage{enumitem}

\usepackage{appendix}
\usepackage{balance}
\usepackage{colortbl}
\usepackage{amsmath}
\usepackage{booktabs}
\usepackage{tabularx}
\usepackage{multirow}
\usepackage{caption}
\usepackage{subcaption}
\usepackage{adjustbox}
\usepackage{lmodern}
\usepackage{tikz}
\usepackage{amssymb}
\usepackage{pifont}

\newcommand\mybox[2][]{\tikz[overlay]\node[fill=blue!20,inner sep=2pt, anchor=text, rectangle, rounded corners=1mm,#1] {#2};\phantom{#2}}

\newcommand{\bbox}[1]{\mybox[fill=blue!20]{#1}}
\newcommand{\rbox}[1]{\mybox[fill=red!20]{#1}}

\newcommand{\tsquare}{\ensuremath{\square}}

\newcommand{\redcmark}{\rbox{\ding{51}}}%
\newcommand{\redxmark}{\cellcolor{red!20}}%
\newcommand{\bluecmark}{\bbox{\ding{51}}}%
\newcommand{\bluexmark}{\cellcolor{blue!20}}%

\setitemize{noitemsep}
\setlist{topsep=0pt, leftmargin=*}

\usepackage[font=small]{caption}

\newcommand{\uls}{\begin{itemize}[leftmargin=*]}
\newcommand{\ule}{\end{itemize}}
\newcommand{\ols}{\begin{enumerate}[leftmargin=*]}
\newcommand{\ole}{\end{enumerate}}
\newcommand{\li}{\item}
\newcommand{\nv}{\cellcolor{lightgray}}

\usepackage{tikz}
\usepackage{bm}

\newcommand{\locals}{MRSQ}
\newcommand{\stands}{SRMQ}
\newcommand{\globs}{MRMQ}

\newcommand{\local}{MRSQ-PP}
\newcommand{\stand}{SRMQ-PP}
\newcommand{\glob}{MRMQ-PP}
\newcommand{\pstand}{$\mathcal{P}_{\text{\stands}}$}
\newcommand{\plocal}{$\mathcal{P}_{\text{\locals}}$}
\newcommand{\pglob}{$\mathcal{P}_{\text{\globs}}$}

\newcommand{\para}[1]{\paragraph{\textnormal{\textbf{#1}.}}}

\usepackage{marginnote}

\begin{document}
%




\title{Breaking Flat: A Generalised Query Performance Prediction Evaluation Framework}

\titlerunning{Multi-Query Multi-Ranker QPP Evaluation}
%

\author{
Payel Santra\inst{1}\orcidlink{0009-0005-5721-248X} \and
Partha Basuchowdhuri\inst{1}\orcidlink{0000-0003-1588-4665} \and 
Debasis Ganguly\inst{2}\orcidlink{0000-0001-7655-7591}
}

\authorrunning{Santra et al.}

\institute{
Indian Association for the Cultivation of Science, Kolkata, India
\\
\email{payel.iacs@gmail.com, partha.basuchowdhuri@iacs.res.in}
\and
University of Glasgow, Glasgow, UK
\\
\email{debasis.ganguly@glasgow.ac.uk}
}

%
\maketitle              

\begin{abstract}

The traditional use-case of query performance prediction (QPP) is to identify which queries perform well and which perform poorly for a given ranking model. A more fine-grained—and arguably more challenging—extension of this task is to determine which ranking models are most effective for a given query. In this work, we generalize the QPP task and its evaluation into three settings: (i) \textbf{Single-Ranker Multi-Query} (\stand), corresponding to the standard use-case; (ii) \textbf{Multi-Ranker Single-Query} (\local), which evaluates a QPP model’s ability to select the most effective ranker for a query; and (iii) \textbf{Multi-Ranker Multi-Query} (\glob), which considers predictions jointly across all query–ranker pairs. Our results show that (a) the relative effectiveness of QPP models varies substantially across tasks (\stand~vs.~\local), and (b) predicting the best ranker for a query is considerably more difficult than predicting the relative difficulty of queries for a given ranker.

%
%
%


\end{abstract}

\section{Introduction}
The standard use case of query performance prediction (QPP) \cite{NQC,uef_kurland_sigir10,DBLP:conf/sigir/MengAAR23,arabzadeh2021bert} is to estimate which queries are likely to perform well for a given ranking model and which are not (see Figure~\ref{fig:eval_stand}). Since such predictions of relative query difficulty are made with respect to a single ranker at a time, the downstream utility of a standard QPP workflow is largely restricted to determining whether that ranker is effective for a query~\cite{DBLP:conf/ecir/DattaGMG24}.
A more practical downstream application of QPP arises in a mixture-of-experts (MoE) setting, where queries could be dynamically ``routed'' to the ranking model most likely to be effective for them. However, the prevailing formulation of the QPP task and its evaluation framework does not provide a means of benchmarking predictors in such MoE retrieval scenarios.

\input{fig_def/fig_evaluation}


In this paper, we move beyond the prevailing \textbf{single-ranker based ``flatlander'' view of QPP} by introducing an \textbf{additional evaluation dimension corresponding to ranking models}.
Not only does this allow to generalise QPP evaluation across multiple queries and ranking models, but it also allows three different ways to conduct the evaluation, each suitable for a characteristically different downstream use case, as shown in Figure \ref{fig:eval_indiv}. This proposed generalised two dimensional QPP evaluation requires to compute a matrix of QPP estimates of the form $\bm{\phi} \in \mathbb{R}^{|Q|\times |\Theta|}$, where $\Theta$ denotes a set of available ranking models and $Q$ denotes a set of queries, and $\phi(q_i, \theta_j)$ denotes a QPP estimate for the ranked list of documents retrieved for the $i^{\text{th}}$ query with the $j^{\text{th}}$ ranking model.

As shown in Figure~\ref{fig:eval_stand}, the first scenario---Single-Ranker Multi-Query (\stands) performance prediction---corresponds to the standard QPP use case, i.e., estimating which queries are likely to perform well for a given ranker. The effectiveness of a QPP model in this setting is determined by how strongly its predictions correlate with the true effectiveness values (e.g., AP or another target IR metric). Higher correlations indicate better predictive quality. For each ranker, correlation values are computed across queries and then averaged, yielding a single score that reflects the overall QPP effectiveness across a set of ranking models (see Figure~\ref{fig:eval_stand}).

The second scenario---Multi-Ranker Single-Query (\locals) performance prediction ---is illustrated in Figure~\ref{fig:eval_local}. Here, the task shifts from predicting relative performance across queries to the more fine-grained challenge of predicting the relative effectiveness of different ranking models for a single query. In this case, a QPP model should be able to discriminate between effective and ineffective rankers on a per-query basis. Unlike the column-wise evaluation in Figure~\ref{fig:eval_stand}, Figure~\ref{fig:eval_local} adopts a row-wise perspective, where correlations are computed across rankers for each query and then averaged over all queries. This procedure provides an estimate of a QPP model’s ability to rank available retrieval models according to their effectiveness for a given query.

Importantly, an effective performance in distinguishing between queries does not necessarily imply an effective performance in distinguishing between rankers, and vice versa. Our work in this paper of generalizing the QPP evaluation framework, allows provision for a thorough empirical evaluation of this question. 

The third scenario --- Multi-Ranker Multi-Query (\globs) performance prediction --- as illustrated in Figure \ref{fig:eval_global} denotes a combination of the \stands~and \locals~setups, where a QPP model is considered effective if it can perform well in both distinguishing good queries from bad ones, and also distinguishing good rankers from bad ones. The QPP effectiveness in this case is computed by a correlation over the entire set of $|Q| |\Theta|=nm$ query-ranker pairs.

\para{Our Contributions}
In summary, following are our main contributions.
\uls 
\li Generalisation of the one dimensional QPP evaluation into two dimensions.
\item A detailed analysis of the performance of existing QPP models for the three different tasks - \stands~(good vs. bad queries), \locals~(good vs. bad rankers), and \globs~(good queries and good rankers vs. bad queries and bad rankers).
\li Findings that score-based QPP approaches perform well for \stand, whereas content-based ones are more effective for \local~and \glob. We make the code available for research purposes\footnote{ \url{https://github.com/gdebasis/eval-precise-qpp}}.


\ule 
%
\vspace{10pt}
Notably, this paper focuses on examining how the relative performance of existing QPP approaches changes under our proposed generalized QPP evaluation framework. The development of novel QPP models for the more fine-grained task of predicting performance across different rankers for the same query is left for future work.

\section{Related Work}

Query Performance Prediction (QPP) has long been studied as the task of estimating retrieval effectiveness without access to relevance labels. Early work explored statistical approaches on retrieval scores \cite{NQC,cummins2011improved,zhou2007query,tao2014query}, robustness of document list perturbations based predictors \cite{uef_kurland_sigir10,roitman2017robust}, and learning to rank features inspired predictors \cite{Chifu2018QueryPPA}.
More recent studies explored supervised and neural approaches for QPP by considering both the content and the score features \cite{DBLP:conf/wsdm/DattaGGM22,10.1145/3209978.3210041,arabzadeh2021bert}.


Another line of QPP research has investigated issues of reproducibility, metric sensitivity, and evaluation stability~\cite{Ganguly2022AnAOA,Meng2024QueryPPA,robustness}. Recent studies have further examined the reproducibility of evaluation methodologies by applying ANOVA~\cite{st1989analysis}, thereby improving statistical reliability and interpretability under the traditional single-ranker setting~\cite{Faggioli2022sMAREANA,Faggioli2023QueryPPA}. While QPP evaluation has conventionally relied on holistic correlation measures (e.g., Kendall’s $\tau$) computed over a set of queries, \cite{Faggioli2022sMAREANA} introduced the sMARE (scaled Mean Absolute Ranking Error) metric, which enables the estimation of per-query relative rank shifts in predicted QPP scores. However, unlike our per-query measure in the \local~setting, which can be computed even for a single query, sMARE still requires multiple queries for evaluation.
%
%
Somewhat similar to our idea of applying QPP to identify the most suitable ranker for a given query, the work in~\cite{khramtsova2024leveraging} employs QPP to predict the document collection on which a particular ranking model is likely to be most effective. In analogous terms, their additional axis of evaluation considers a set of document collections rather than a set of rankers as in our work.

%
%
%

Similar to our idea of evaluating QPP systems under an additional evaluation dimension, existing research has evaluated systems by considering aspects of performance measures other than relevance, such as gains in a user's cognitive state of knowledge \cite{search-as-learning}, visual complexity \cite{visualcomplexity}, explainability \cite{dg_neuralRankers},
exposure fairness \cite{retr-analysis-cikm2014,singh2018fairness}, and
consistency \cite{consistency-sigir17,consistency-cikm22}. Additionally,
Kasai et al.~\cite{Kasai2021BidimensionalLGA} proposed bidimensional leaderboards that jointly consider both generation and evaluation aspects of language models, and likewise, Massiah et al.~\cite{massiah2023reliability} examined conversational agent evaluation along multiple dimensions, highlighting the importance of reliability and variability in user feedback. Santra et al.~\cite{santra2025hf} proposed a hierarchical fusion-based framework that combines relevant contexts from multiple rankers and corpora in RAG context.

\section{Generalized QPP: Task and Evaluation} \label{sec:method}

In Section \ref{ss:standard}, we first formally define the standard QPP task as a Single-Ranker Multi-Query Performance Prediction (\stand), following which we generalize this to Multi-Ranker Single-Query (\local) and Multi-Ranker Multi-Query (\glob) setups, respectively, in Sections \ref{ss:local} and \ref{ss:global}.

\subsection{Single-Ranker Multi-Query Performance Prediction} \label{ss:standard}


The standard QPP task involves predicting the quality of the top-$k$ documents $L_k(q, \theta)$ retrieved by a ranking model $\theta$ in response to a query, \textbf{relative to the performance measure observed for another query}. Standard QPP is typically formalized as a function of the form
$\phi(q, L_k(q, \theta)) \mapsto \mathbb{R}$
which outputs a predicted performance value given an input query $q$. 

The evaluation of a QPP model $\phi$ \textbf{must be executed on a set of queries}, by computing the correlation between the predicted and the corresponding ground-truth performance measures $\mu$ (e.g., nDCG@10, AP@50). It is to be noted that, the correlation measures, e.g., Kendall’s $\tau$ are undefined for singleton sets of paired values.
Even more recent measures, such as sMARE~\cite{Faggioli2022sMAREANA}, which provide per-query insights into QPP effectiveness, rely on computing rank shifts across queries and therefore still require evaluation over multiple queries.

This implies that a predictor is considered effective if it assigns higher estimates to a query, leading to a better retrieval quality than another using the same retrieval method. In other words, an effective model is capable of distinguishing between different queries, which indicates that both the prediction and its evaluation rely on \textbf{relative performance comparisons across multiple queries}. Formally,
\begin{equation}
\mathcal{P}_{\text{SRMQ}}(Q, \theta, \phi, k, \mu) = \chi\left(\bigcup_{q \in Q} \mu(L_k(q, \theta)), \bigcup_{q \in Q} \phi(L_k(q, \theta))\right), \label{eq:standard}
\end{equation}
where $\chi$ is a correlation measure (e.g., Kendall's $\tau$) between two ordered sets (lists) of values -- in this case the correlation between the per-query IR performance measures $\mu$, and the QPP estimates for the corresponding queries.   

Although existing research in QPP reports the performance separately for each ranker used in the experiments, it is also possible to compare QPP approaches by their average performance across a range of different rankers, i.e.,
\begin{equation}
\mathcal{P}_{\text{SRMQ}}(Q, \phi, k, \mu) = \frac{1}{|\Theta|}\sum_{\theta \in \Theta}\mathcal{P}_{\text{SRMQ}}(Q, \theta, \phi, k, \mu),     \label{eq:standard_ext}
\end{equation}
where $\Theta$ denotes a set of rankers used to compute the overall performance of a predictor. The schematic workflow of \stand~setup is depicted in Figure \ref{fig:eval_stand}.

\subsection{Multi-Ranker Single-Query Performance Prediction} \label{ss:local}
A pair of queries with largely different information needs (as is usually the case for constituent queries of a benchmark IR dataset, e.g., TREC DL) can lead to top-retrieved sets of documents with little or no overlap. As a result, the primary signals on which the QPP models rely, e.g., the score distribution of unsupervised QPP approaches, and the topical interaction between query-document pairs of supervised ones, are expected to be substantially different. This makes the task relatively easy to distinguish between easy and difficult queries.


We consider a more fine-grained variant of the standard QPP task, which, specifically speaking, involves predicting which ranking model is likely to yield a better retrieval performance than another model for the same query. In contrast to the standard task, this multi-ranker setup for QPP is likely to pose significant challenges for standard QPP approaches, mainly due to the fact that there is now significant topical overlap in the set of top-retrieved documents.
This means that there may be no significant differences in the relative score distributions or between the topical interaction between query-document pairs, respectively, for unsupervised and supervised QPP approaches to work effectively.


Instead of fixing a ranking model and then predicting the relative difficulty of queries with respect to that model, we propose an alternative perspective: fix a query first and \textbf{predict the relative effectiveness of different rankers with respect to that query}.
%
%
More formally, the multi-ranker single-query performance prediction (MRSQ) approach is evaluated as
\begin{equation}
\mathcal{P}_{\text{MRSQ}}(q, \Theta, \phi, k, \mu) = \chi\!\left(\bigcup_{\theta \in \Theta} \mu(L_k(q, \theta)), \bigcup_{\theta \in \Theta} \phi(L_k(q, \theta))\right),    
\end{equation}
where $q$ denotes an individual query. The measure $\mathcal{P}_{\text{MRSQ}}(q, \theta, \phi, k, \mu)$ quantifies how effectively a QPP model can predict the relative ranking of retrieval methods ($\theta \in \Theta$) for a given query--that is, which ranker is expected to perform best, which is second-best, and so on. This per-query effectiveness measure can then be aggregated across a set of benchmark queries as
\begin{equation}
\mathcal{P}_{\text{MRSQ}}(\Theta, \phi, k, \mu) = \frac{1}{|Q|} \sum_{q \in Q} \mathcal{P}_{\text{MRSQ}}(q, \Theta, \phi, k, \mu). \label{eq:local}
\end{equation}
Notably, the evaluation framework defined in Equation~\ref{eq:local} (schematic depiction in Figure \ref{fig:eval_local}) enables the computation of per-query QPP effectiveness and is therefore amenable to statistical significance testing, such as a paired $t$-test.

\subsection{Multi-Ranker Multi-Query Performance Prediction} \label{ss:global}

The underlying task in \stand~consists of predicting the relative retrieval effectiveness of queries for a given ranker, whereas the task in \local~focuses on predicting the relative effectiveness of a set of rankers for a particular query. In both cases, evaluation typically requires computing average correlation values across queries or rankers, respectively.  

These two tasks, i) predicting the relative effectiveness of queries and ii) predicting the relative effectiveness of rankers, can be unified into a joint objective of predicting the effectiveness of each query-ranker pair. Under this formulation, the evaluation of the Multi-Ranker Multi-Query Performance Prediction (\glob) task reduces to computing a single correlation measure over the set of all query–ranker pairs, thereby eliminating the need for averaging as in Equations \ref{eq:standard} and \ref{eq:local}. More formally,
\begin{equation}
\mathcal{P}_{\text{MRMQ}}(\phi, k, \mu) = \chi\!\left(\bigcup_{\theta \in \Theta} \bigcup_{q \in Q} \mu(L_k(q, \theta)), \bigcup_{\theta \in \Theta} \bigcup_{q \in Q} \phi(L_k(q, \theta))\right), \label{eq:global}  
\end{equation}
where, different from Equations \ref{eq:standard} and \ref{eq:local}, the effectiveness of a QPP model $\phi$ is obtained by a correlation between true and predicted retrieval performance measures computed over each ranker-query pair, i.e., over a total of $|\Theta||Q|=nm$ pairs, as schematically depicted in Figure \ref{fig:eval_global}.

\section{Experiment Setup}

\subsection{Research Questions}
The first research question we address concerns the standard QPP setting, with the key distinction that, instead of reporting QPP effectiveness for a small number of retrieval models individually, we report the average correlation (Equation~\ref{eq:standard}) across eight IR models (see Section~\ref{ss:rankers}). The objective is to enable a thorough comparison of existing QPP models (see Section~\ref{ss:qpp-methods}), considering not only their effectiveness for individual rankers but also their average performance across multiple rankers. Stated explicitly,
\uls
\li \textbf{RQ1}: What are the best performing QPP models in the generalized \stand~setting, i.e., when performance measures are averaged across several ranking models of different characteristics (Equation \ref{eq:standard})?
\ule
Our next research question is related to the proposed generalization of a standard QPP task into its more fine-grained version -- \local, where a QPP model now needs to predict the best ranker available for each query.
\uls
\li \textbf{RQ2}: Do the best performing QPP models also continue to work most effectively in the \local~setting, i.e., predicting which ranker is likely to yield the best retrieval results for a query?
\ule
%
%
%
%
Since the \local~setup yields per-query QPP effectiveness, this allows provision to conduct a statistical significance testing between different QPP models, which is not possible in the standard QPP evaluation workflow. Explicitly,
\uls
\li \textbf{RQ3}: Are the relative performance differences between standard QPP models statistically significant as per the \local~setup? 
\ule
Finally, we investigate if the evaluation of a more generalized QPP task to this 2D setup, \glob~(i.e., ranker and query instead of anyone alone) should be evaluated by a combined correlation (Equation \ref{eq:global}) as opposed to a combination of two separate evaluation measures for the \stand~and \local~setups.
\uls
\li \textbf{RQ4}: What are the benefits of computing the correlation over all query-ranker pairs -- $\mathcal{P}_{\text{MRMQ}}(\phi, k, \mu)$ of Equation \ref{eq:global}) -- vs. the harmonic mean (F1-score like) based combination of the average values of the per-ranker and per-query correlations (Equations \ref{eq:standard} and \ref{eq:local})?
\ule 

%


\subsection{IR Models Investigated} \label{ss:rankers}

Since one of the objectives of our proposed generalized QPP evaluation framework is to examine how effectively existing QPP approaches predict the performance of ranking models, we employ a diverse set of ranking models with different characteristics to enable a reasonably comprehensive evaluation. The broad categories of ranking models that we employ are: sparse, learned sparse, dense (bi-encoders), and retrieve-and-rerank (cross-encoders). 


\para{(Learned) Sparse Retrieval}
\uls
\li \textbf{BM25}~\cite{10.1561/1500000019}: A classical IR model that relies on exact term matching (via a sparse index). The term weight is a variant of tf-idf with a factor $b$ for document length normalisation.

\li \textbf{RM3}~\cite{10.1145/3471158.3472261}: A pseudo-relevance feedback (PRF) model that enriches the initial query with terms from the top-retrieved documents that frequently co-occur with the original query terms.

\li \textbf{SPLADE}~\cite{formal2021splade}: A learned sparse retrieval model that expands queries with contextual terms under sparsity regularization, reducing vocabulary mismatch while retaining compatibility with inverted indexes.
\ule



\para{Dense Retrieval}
\uls
\li \textbf{E5}~\cite{wang2022text}: A dense retrieval model trained with weakly-supervised contrastive learning on large-scale curated text pairs.
\li \textbf{ColBERT}~\cite{10.1145/3397271.3401075}: A late interaction model that captures fine-grained token-level interactions.

\li \textbf{ColBERT-PRF}~\cite{wang2023colbert}: A dense vector space based PRF model that first clusters token embeddings from top-ranked documents and appends the most representative ones to the query. 
\ule


\para{Retrieve-and-Rerank}
\uls
\li \textbf{BM25>>MonoT5}~\cite{DBLP:journals/corr/abs-2101-05667}, a retrieve-and-rerank pipeline, which uses BM25 for retrieving the top-100 documents which are then reranked by MonoT5, a T5-based cross-encoder. 

\li \textbf{BM25>>ColBERT-PRF reranker}~\cite{wang2023colbert}: Uses BM25 as the initial retriever and then employs local co-occurrence statistics to obtain a set of candidate token embeddings for expanding the original query and reranking.

\ule

\subsection{QPP Models Investigated} \label{ss:qpp-methods}

We employ a wide range of diverse QPP models to investigate their performance under the proposed generalized evaluation framework.
The hyper-parameters of each QPP model, e.g., the cut-off rank for NQC, the fraction $x$ for $n(\sigma_{x\%})$ etc., were set to the optimal values reported in the corresponding papers.

\para{Unsupervised QPP Models}
\uls 
\li \textbf{NQC}~\cite{NQC}: Computes the standard deviation of the scores of the top-retrieved documents with the assumption that a skewed distribution (higher variance) of these scores likely indicates better retrieval.


\li \textbf{WIG} \cite{zhou2007query}: Measures the
aggregated value of the information gain of each top-retrieved document with respect to the collection.

\li $\sigma_\text{max}$~\cite{perez2010standard}: Computes the maximum of the standard deviation of the retrieval scores over an iteratively increasing set of top-retrieved documents.



\li $n(\sigma_{x\%})$~\cite{cummins2011improved}: Computes the standard deviation of the retrieval scores for the top-ranked documents whose scores fall within a fraction $x \in [0,1]$ of the highest-ranked document's score.





\li \textbf{SMV}~\cite{tao2014query}: A modified version of NQC that, instead of relying on score variance, computes a weighted sum of the ratios between individual retrieval scores and the mean score.




\li \textbf{UEF}~\cite{uef_kurland_sigir10}: This method computes a weighted average of QPP estimates (obtained from a base QPP model, such as NQC) over sublists sampled from the top-retrieved document list. A higher degree of perturbation in a sampled top list after the feedback operation indicates a greater likelihood that the retrieval effectiveness of the original list was poor, thereby suggesting that lower confidence should be assigned to the QPP estimate for such a query.
In our experiments, we set the base QPP estimator of UEF as NQC following the common practice in existing works~\cite{arabzadeh2023noisy,arabzadeh2021bert,datta2022pointwise}.

\li \textbf{RSD}~\cite{roitman2017robust}: A simplified version of UEF where QPP estimations are averaged over sublists randomly sampled from the top-retrieved list of documents.
%


\li \textbf{QPP-PRP}~\cite{singh2023unsupervised}: Measures the consistency of an input ranked list with the pairwise rank preferences derived from an auxiliary model, such as Duo-T5~\cite{DBLP:journals/corr/abs-2101-05667}. For our experiments, we employ a simplified variant of~\cite{singh2023unsupervised}, in which rather than computing an odds ratio over the full matrix of pairwise preferences, the input list is partitioned into top and bottom segments.



\li \textbf{SCNQC}~\cite{roitman2019normalized} is a generalization of the NQC predictor that introduces three hyper-parameters to control score normalization, deviation weighting, and idf-based scaling.


\li \textbf{QV-NQC}~\cite{Oleg_2019}: Combines the QPP estimate of the original query with QPP estimates computed over alternate formulations of similar information needs. Instead of using generated query variants \cite{datta2022relative}, given a query, we use the MS MARCO training set to retrieve a candidate set of similar information needs.

\li \textbf{DM} (Discounted Matryoshka)~\cite{10.1145/3539618.3591625}: Here the QPP estimate is the diameter of the embedded representations of the top-ranked documents -- a smaller value indicating lower diversity, and likely a better retrieval performance.

\ule 

\para{Supervised QPP Models}
\uls
\li \textbf{NQA-QPP}~\cite{hashemi2019performance}: Uses a combination of features (e.g., scores) and embedded vectors to train a regression model on retrieval quality.

\input{fig_def/main_table}
\input{fig_def/main_fig}

\li \textbf{BERTQPP}~\cite{arabzadeh2021bert}: This approach trains a regression head to learn an IR evaluation metric on a BERT-based cross-encoder that jointly encodes the content of a query and its top-retrieved document.

\ule


\subsection{IR Dataset and Evaluation Measures} \label{ss:datasets}
We experiment with the standard IR benchmark datasets, namely
TREC DL'19 and TREC DL'20, comprising a total of 97 queries~\cite{craswell2025overview}, as is commonly used in the QPP literature \cite{arabzadeh2021bert,DBLP:conf/wsdm/DattaGGM22,datta2022relative}. For a focused analysis, we conduct all our experiments on the combined DL'19 and '20 topic sets, which
reduces the total number of results to report in comparison to reporting results separately for the two sets.
The underlying document collection for these queries is the MS MARCO passage collection comprising over 8.8M document passages \cite{msmarco-data}. For training the supervised QPP approaches used in our experiments (more details in Section \ref{ss:qpp-methods}), we use the MS MARCO training dataset, which consists of approximately 500K queries paired with relevance judgments. 

As target IR evaluation metrics to evaluate the QPP models (i.e., the function $\mu$ in Equations \ref{eq:standard}, \ref{eq:local} and \ref{eq:global}), we employ AP@50 and nDCG@10. While the former balances both precision and recall, the latter represents a precision-only metric more suitable for web search.
%
Additionally, as a concrete realisation of the correlation measure $\chi$ of Equations \ref{eq:standard}, \ref{eq:local} and \ref{eq:global}), we use the Kendall’s $\tau$ function.


\subsection{Implementation Details} \label{ss:implementation}
For all unsupervised QPP methods, we use the hyperparameters
mentioned in previous studies and for different predictors we set different values of $k$ . Following \cite{zhou2007query} $k$ is set to 5 for WIG. As suggested by \cite{NQC,tao2014query}, we set $k = 100$ for NQC and SMV. Following \cite{cummins2011improved}, we set $\beta = 0.5$ for $n(\sigma_{x\%})$. The $\sigma_{\text{max}}$ predictor does not require any hyperparameters. For QPP-PRP, we set $p = 0.2$. For DM, we set $k = 5$ and compute dense vectors using the \url{sentence-transformers/all-MiniLM-L6-v2} model.

\section{Results and Discussions}

\input{table/correlations}
\input{table/significant_testing}

\para{Main findings}


For \textbf{RQ1}, the Table~\ref{fig:main_table_final} shows that, in terms of \textbf{average performance across multiple rankers} (\stands~column), \textbf{unsupervised methods} such as NQC and RSD \textbf{outperform supervised approaches} by substantial margins. This finding contrasts with recent reports~\cite{arabzadeh2021bert,datta2022pointwise,hashemi2019performance,meng2025query}, and highlights that when QPP evaluation incorporates an additional ranker dimension, classical retrieval-score-based unsupervised methods (e.g., NQC, RSD, UEF) consistently outperform modern content-based supervised approaches (e.g., NQA-QPP, BERT-QPP). In addition to the average results reported in the Table~\ref{fig:main_table_final}, the Figures \ref{fig:rc_box_ap} and \ref{fig:qc_box_ap} present more detailed per-ranker and per-query correlations.
%
\\
While unsupervised QPP approaches are generally more effective at distinguishing easy from difficult queries for a given ranker, the \locals~column in the Table~\ref{fig:main_table_final} reveals the opposite trend for \textbf{RQ2}, which concerns predicting the most effective ranking model for a given query. Specifically, we observe that \textbf{content-based approaches}---namely the supervised BERT-QPP and the embedding-based unsupervised DM---\textbf{perform substantially better} than traditional unsupervised methods in \textbf{discriminating effective rankers from ineffective ones} with respect to AP@50. Furthermore, and as related to \textbf{RQ3}, many of these improvements are statistically significant, as reported in Table~\ref{tab:ttest_loca}. In contrast, when effectiveness is measured using nDCG@10 (see the \locals~column in Table~\ref{fig:main_table_final}), the differences across QPP models are negligible.

In relation to \textbf{RQ4}, i.e., the overall two dimensional QPP evaluation across a set of queries and rankers, it is observed from the \globs~column of the Table \ref{fig:main_table_final} that similar to \locals, BERTQPP and DM continue to be the best performing QPP methods. As overall observations, we note that the task of predicting the best performing queries for a ranker (\stand) is much easier than the task of predicting the best performing ranker to which a query should be routed (\local), as can be seen from the substantially larger average $\tau$ values in the \stands~column as compared to the \locals~column. 

\para{Correlation between \stand, {\local}, and \glob}
From Table \ref{tab:correlations} it is evident that the two tasks \stand~and \local~yield substantially different relative effectiveness of QPP models with AP@50 as the target IR measure as can be seen from the low correlation value (this is also illustrated later in Table \ref{tab:ttests_ap}). However, with a more precision-oriented measure, i.e., nDCG@10, the differences between the relative performance of the QPP models decrease, as can be seen from the much higher correlation between \stand~and \local. In relation to RQ4, we observe that there is small correlation between a global correlation across ranker-query pairs ($\mathcal{P}_{\text{\globs}}$) and the simple F1-score like combination of the two different complementary measures $\mathcal{P}_{\text{\stands}}$ and $\mathcal{P}_{\text{\locals}}$. However, due to the higher standard deviations of $\mathcal{P}_{\text{\globs}}$ the global correlation can be considered to be a better discriminator between QPP models than the F1-score combination.

\para{Pairwise Significance}

In Section \ref{sec:method} we stated that with the additional dimension of ranking model included for QPP evaluation, it is now possible to perform a significance testing (paired t-test) between the average correlation values computed either across queries (Equation \ref{eq:standard}) or across rankers (Equation \ref{eq:local}). To address RQ3, Table \ref{tab:ttests_ap} provides further insight into the statistical significance of pairwise comparisons between different QPP models for the \stands~and \locals~tasks. By comparing the upper triangles of Figures \ref{tab:ttest_stand} and \ref{tab:ttest_loca} we observe a large number of changes in cell colours (\bbox{\tsquare} and \rbox{\tsquare}), indicating that \textbf{the relative performance of QPP models changes across these two tasks} (predicting the best query for a ranker vs. the best ranker for a query). It is also interesting to see that the more difficult task, namely \local, leads to a smaller number of significant differences.

\section{Conclusions and Future work}
In this paper, we extended the conventional one-dimensional view of query performance prediction (QPP) into a two-dimensional framework that jointly considers queries and ranking models. Our findings demonstrate that the task of distinguishing effective from ineffective rankers for a particular query (\local) is substantially more challenging than the traditional task of distinguishing easy from difficult queries for a particular ranker (\stand). This highlights an important limitation of existing QPP methodologies when applied in more fine-grained evaluation scenarios.

In future, we plan to develop novel QPP approaches for addressing this fine-grained task of predicting the best ranker on a per-query basis. Moreover, similar to \cite{DBLP:conf/ecir/DattaGMG24}, we also plan to apply the QPP estimates for developing an adaptive ranking pipeline.

\subsubsection*{Disclosure of Interests}
We declare that we don't have any competing interests.

\bibliographystyle{splncs04}
\bibliography{references.bib}
\end{document}

%% file: fig_def/fig_evaluation.tex
\begin{figure}[t]
\centering
\begin{subfigure}[b]{0.38\columnwidth}
\centering
\includegraphics[width=1\textwidth]{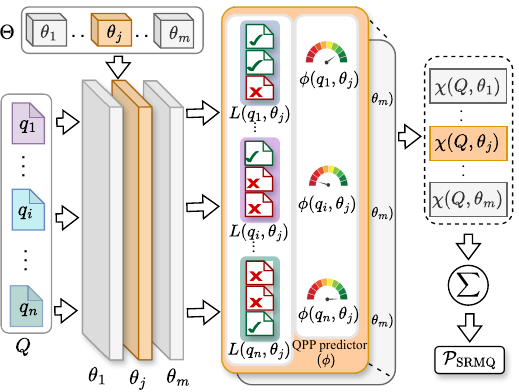}
\caption{\stand}
\label{fig:eval_stand}
\end{subfigure}
\quad
\begin{subfigure}[b]{0.4\columnwidth}
\centering
\includegraphics[width=1\textwidth]{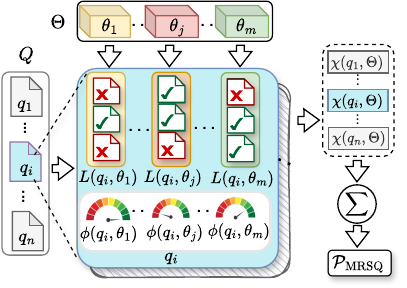}
\caption{\local}
\label{fig:eval_local}
\end{subfigure}
\quad
\begin{subfigure}[b]{0.85\columnwidth}
\centering
\includegraphics[width=1\textwidth]{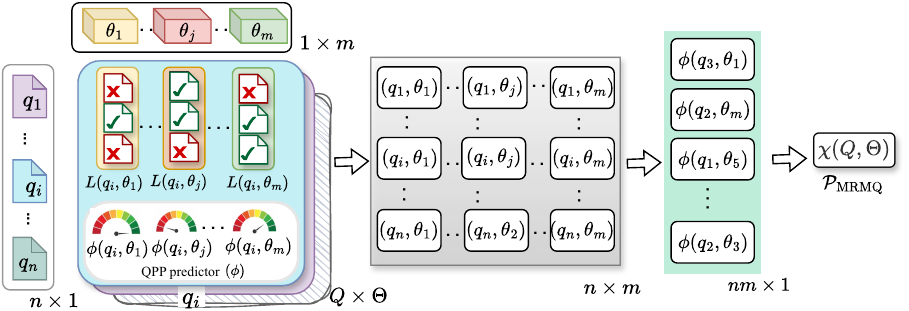}
\caption{\glob}
\label{fig:eval_global}
\end{subfigure}
\caption{A generalised QPP evaluation framework with an additional dimension of ranking models -- \textbf{a)}:
each slice represents per-ranker QPP estimates across a set of queries and their correlations with the corresponding ground-truth retrieval quality (Equation \ref{eq:standard}, \ref{eq:standard_ext});
\textbf{b)}: each slice represents per-query predictions of rankers' performance and their correlations with the corresponding ground-truth retrieval quality for each IR model (see Equation \ref{eq:local});
\textbf{c)}: computes a correlation between $|\Theta| |Q|=nm$ predictions and retrieval quality for each query–ranker pair (Equation \ref{eq:global}).
}
\label{fig:eval_indiv}
\end{figure}


%% file: fig_def/main_table.tex
\begin{table*}[t]
\centering
\centering
\small
\begin{adjustbox}{width=0.7\linewidth}
\begin{tabular}{@{}lccc ccc@{}}
\toprule
 & \multicolumn{3}{c}{AP@50} & \multicolumn{3}{c}{nDCG@10} \\
\cmidrule(r){2-4}
\cmidrule(r){5-7}
$\phi$ & \pstand & \plocal & \pglob & \pstand & \plocal & \pglob \\
\midrule
NQC           &  \underline{\textbf{.381}} & .085 & .109 &  \underline{.274} & .106 &  .033 \\
WIG           &  \underline{.125} &  .032  & .028 &  \underline{.120} &  .054 & -.005 \\
$\sigma_{\text{max}}$ &  \underline{.277} &  .082  & .079 &  \underline{.184} &  .102 &  .005 \\
$n(\sigma_{x\%})$  &  \underline{.266} &  .091 & .084 &  \underline{.184} &  .103  & .013 \\
SMV           &  \underline{.345} & .082&  .119 &  \underline{.230} &  .103 & .039 \\
UEF           &  \underline{.364} & .082  & .106 &  \underline{.260} &  \textbf{.113} &  .028 \\
RSD           &  \underline{.372} & .087  &  .108 &  \underline{\textbf{.277}} &  .104 &  .036 \\
QPP-PRP       &  \underline{.212} & .107 &  .164 &  \underline{.117} & .037 &  .090 \\
SCNQC         &  \underline{.381} & .085  &  .109 &  \underline{.274} &  .106 & .033 \\
QV-NQC        &  \underline{.312} & .089&  .114 &  \underline{.202} & .103 & .027 \\
DM            &  .174 & .104  &  \underline{.193} &  .055 & .052 &  \underline{.077} \\
NQA-QPP       &  \underline{.167}  & .074 &  .117 &  .034  &  \underline{.109} & .003 \\
BERTQPP       &  \underline{.209} & \textbf{.117} &  \textbf{.207} &  \underline{.147} & .094 &  \textbf{.143} \\
\bottomrule
\end{tabular}
\end{adjustbox}
\caption{Comparison between a total of 13 different QPP methods on a total of 97 queries (TREC-DL’19\&20) for three different task and evaluation setups: \stand, \local, and \glob. Each reported value is a Kendall’s $\tau$ correlation of QPP estimates with two target IR metrics: AP@50 and nDCG@10. Each best-performing QPP model for a task-metric combination (i.e., the highest values along a column) is bold-faced. Similarly, the task for which a QPP model is the most effective (i.e., \stand, \local, or \glob) for a specific target IR metric, is shown by underlining the best values across each row.
}
\label{fig:main_table_final}
\end{table*}

%% file: fig_def/main_fig.tex
\begin{figure}[t]   
\centering
\begin{subfigure}{0.4\linewidth}
  \centering
  \includegraphics[width=\linewidth]{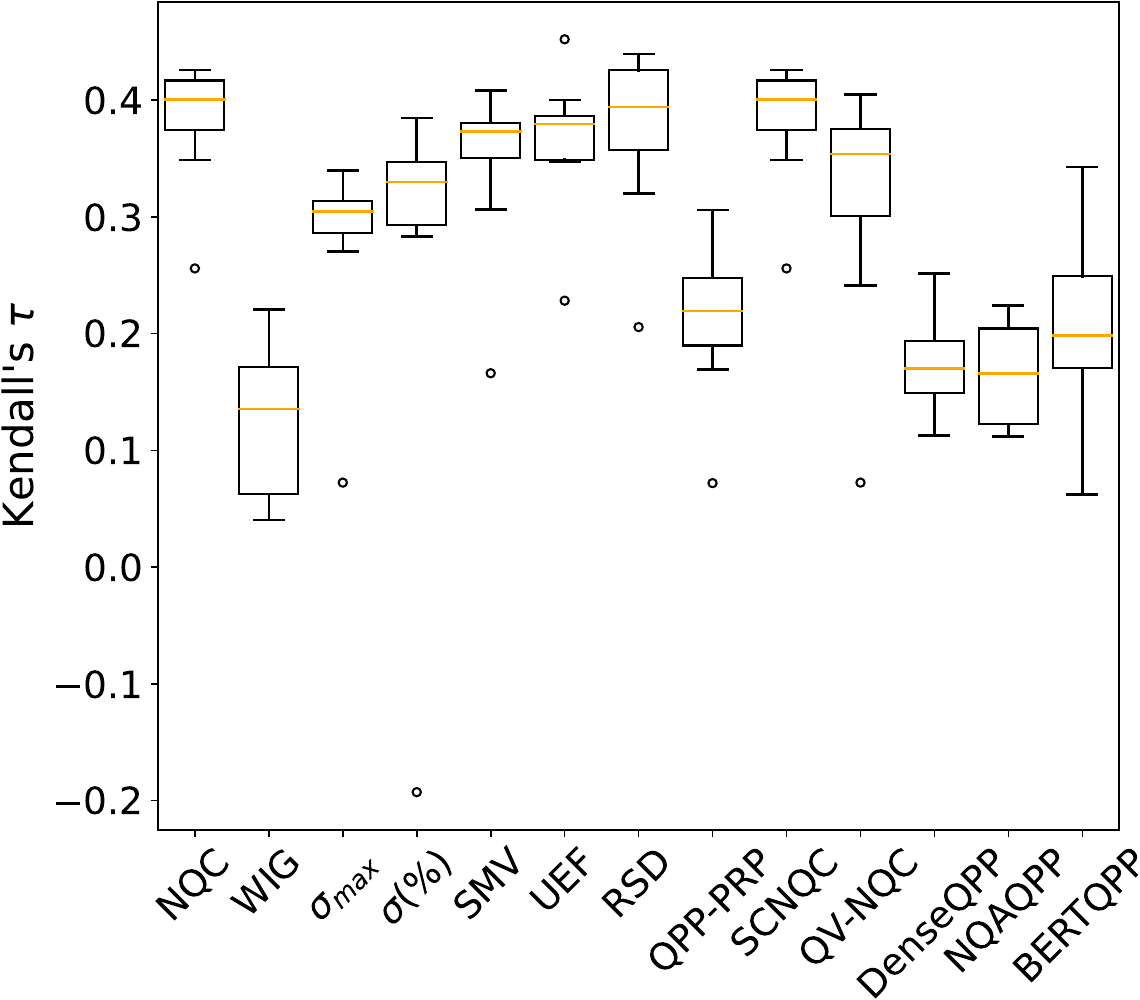}
  \caption{\stand}
  \label{fig:rc_box_ap}
\end{subfigure}
\quad
\begin{subfigure}{0.4\linewidth}
  \centering
  \includegraphics[width=\linewidth]{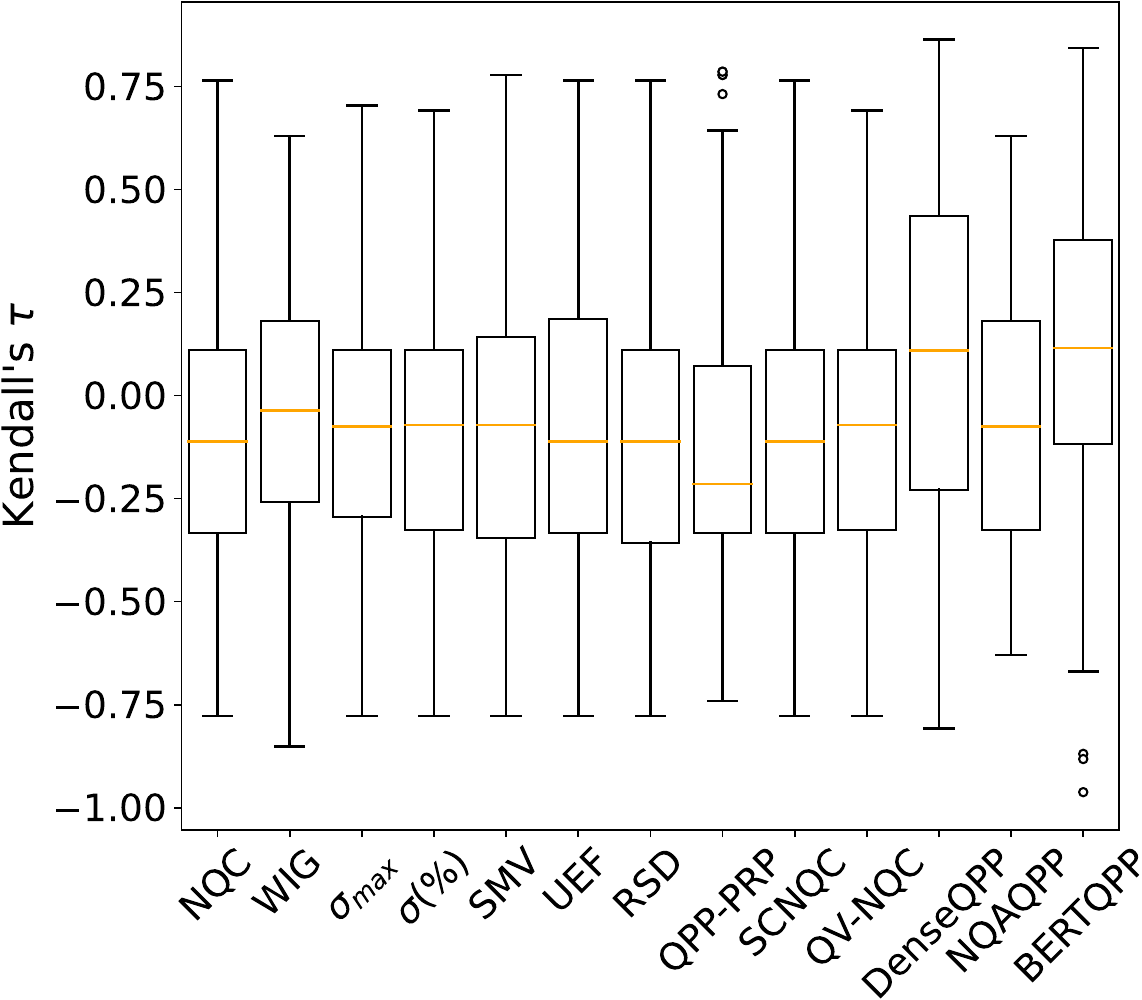}
  \caption{\local}
  \label{fig:qc_box_ap}
\end{subfigure}
\caption{\small Per-ranker and per-query Kendall's $\tau$ values for a) \stands~and b) \locals~settings with AP@50 as the target IR measure.
The variations across queries in \local~are much higher than those across rankers in \stand.
}
\label{fig:table_and_boxplot}
\end{figure}

%% file: table/correlations.tex
\begin{table*}[t]
\begin{subtable}[b]{0.49\textwidth}
\centering
\begin{adjustbox}{width=0.99\columnwidth}
\begin{tabular}{@{}lrrr@{}}
\toprule
\multicolumn{2}{c}{} & \multicolumn{2}{c}{Correlation}  \\
\cmidrule(r){3-4}
\multicolumn{2}{c}{Metrics} & AP@50 & nDCG@10 \\
\midrule
\pstand & \plocal &  \textbf{-.053} & .520 \\
\pstand & \pglob & -.091  & .183 \\
\plocal & \pglob & .450 &\textbf{-.239}  \\
F1 (\pstand, \plocal) & \pglob & .299 & .065 \\
\bottomrule
\end{tabular}
\end{adjustbox}
\caption{Correlations between evaluation metrics}
\label{tab:correlations}
\end{subtable}
\quad
\begin{subtable}[b]{0.49\textwidth}
\centering
\begin{adjustbox}{width=0.8\columnwidth}
\begin{tabular}{@{}lcc@{}}
\toprule
& \multicolumn{2}{c}{Std. Dev ($\sigma$)} \\
\cmidrule(r){2-3} 
Metric & AP@50 & nDCG@10 \\ 
\midrule
F1 ($\mathcal{P}_{\text{\stands}}$, $\mathcal{P}_{\text{\locals}}$) & .0257 & .0411\\
$\mathcal{P}_{\text{\globs}}$ & \textbf{.0473} & \textbf{.0414}\\
\bottomrule
\end{tabular}
\end{adjustbox}
\caption{Variations of the evaluation metrics.}
\label{tab:final_res_ndcg}
\end{subtable}
%
\caption{QPP evaluation measures: 
(a) Low correlations (boldface indicates the lowest value in each column) highlight inherent differences between task-specific evaluation measures.  
(b) High standard deviations (boldface indicates the highest value in each column) across $13$ QPP models suggest that the corresponding measure is more discriminative, i.e., better able to distinguish effective from ineffective predictors.}
%
\label{tab:variances}
\end{table*}

%% file: table/significant_testing.tex
\begin{table*}[t]
\centering
\begin{subtable}{0.37\textwidth}
\centering
\begin{adjustbox}{width=0.99\columnwidth}
\begin{tabular}{l ccc ccc ccc cccc}
 $\phi$ & \rotatebox[origin=c]{90}{NQC}  & \rotatebox[origin=c]{90}{WIG} & \rotatebox[origin=c]{90}{$\sigma_{\max}$} & \rotatebox[origin=c]{90}{$\sigma(\%)$} & \rotatebox[origin=c]{90}{SMV} & \rotatebox[origin=c]{90}{UEF} & \rotatebox[origin=c]{90}{RSD} & \rotatebox[origin=c]{90}{QPP-PRP} & \rotatebox[origin=c]{90}{SCNQC} & \rotatebox[origin=c]{90}{QV-NQC} & \rotatebox[origin=c]{90}{DM} & \rotatebox[origin=c]{90}{NQAQPP} &\rotatebox[origin=c]{90}{BERTQPP} \\
NQC          &  \nv    & \redcmark   & \redcmark   & \redcmark  & \redcmark & \redxmark & \redxmark & \redcmark  & \redxmark & \redxmark & \redcmark & \redcmark& \redcmark\\
WIG           &   \nv    &    \nv     &    \bluecmark  & \bluexmark  &  \bluecmark & \bluecmark & \bluecmark& \bluecmark & \bluecmark &  \bluecmark& \bluecmark&\bluecmark & \bluecmark\\
$\sigma_{\max}$  &    \nv   &     \nv    &   \nv    &  \redxmark & \bluecmark & \bluecmark & \bluecmark& \redcmark & \bluecmark & \bluexmark & \redcmark & \redcmark & \redxmark\\
$\sigma(\%)$   &    \nv   &    \nv     &    \nv   &   \nv & \bluexmark & \bluexmark &  \bluecmark & \redxmark &  \bluecmark & \bluexmark &  \redxmark& \redxmark& \redxmark \\
SMV           &     \nv  &    \nv     &  \nv     &   \nv & \nv & \bluexmark & \bluecmark & \redcmark & \bluecmark & \redxmark & \redcmark & \redcmark & \redcmark \\
UEF          &    \nv   &    \nv     &    \nv   &  \nv  &   \nv& \nv & \bluexmark &  \redcmark & \bluexmark & \redxmark & \redcmark & \redcmark & \redcmark \\
RSD          &   \nv    &    \nv     &     \nv  &  \nv  &   \nv&   \nv & \nv& \redcmark  & \bluexmark & \redxmark  &  \redcmark & \redcmark & \redxmark\\
QPP-PRP     &   \nv    &    \nv     &     \nv  &  \nv  &   \nv&   \nv & \nv & \nv & \bluecmark & \bluecmark & \redcmark& \redcmark & \redcmark \\
SCNQC       &   \nv    &    \nv     &     \nv  &  \nv  &   \nv&   \nv & \nv & \nv  & \nv & \redxmark & \redcmark & \redcmark & \redcmark \\
QV-NQC      &   \nv    &    \nv     &     \nv  &  \nv  &   \nv&   \nv & \nv & \nv & \nv & \nv & \redcmark & \redcmark & \redcmark\\
DM        &   \nv    &    \nv     &     \nv  &  \nv  &   \nv&   \nv & \nv & \nv & \nv & \nv  &\nv & \bluexmark & \redxmark \\
NQAQPP      &   \nv    &    \nv     &     \nv  &  \nv  &   \nv&   \nv & \nv & \nv & \nv & \nv & \nv &\nv & \redxmark\\
BERTQPP     &   \nv    &    \nv     &     \nv  &  \nv  &   \nv&   \nv & \nv & \nv & \nv &  \nv&\nv &\nv & \nv\\
\end{tabular}
\end{adjustbox}
\caption{\stand}
\label{tab:ttest_stand}
\end{subtable}
\quad
\begin{subtable}{0.37\textwidth}
\centering
\begin{adjustbox}{width=0.99\columnwidth}
\begin{tabular}{l ccc ccc ccc cccc}
 $\phi$ & \rotatebox[origin=c]{90}{NQC}  & \rotatebox[origin=c]{90}{WIG} & \rotatebox[origin=c]{90}{$\sigma_{\max}$} & \rotatebox[origin=c]{90}{$\sigma(\%)$} & \rotatebox[origin=c]{90}{SMV} & \rotatebox[origin=c]{90}{UEF} & \rotatebox[origin=c]{90}{RSD} & \rotatebox[origin=c]{90}{QPP-PRP} & \rotatebox[origin=c]{90}{SCNQC} & \rotatebox[origin=c]{90}{QV-NQC} & \rotatebox[origin=c]{90}{DM} & \rotatebox[origin=c]{90}{NQAQPP} &\rotatebox[origin=c]{90}{BERTQPP} \\
NQC          &  \nv    &    \redcmark    &   \redxmark    &  \bluexmark & \redxmark  & \redxmark & \bluexmark & \bluexmark & \redxmark & \bluexmark & \bluecmark  & \redxmark & \bluecmark\\
WIG           &   \nv    &    \nv     & \bluecmark     &   \bluecmark & \bluecmark & \bluecmark & \bluecmark & \bluecmark & \bluecmark & \bluecmark & \bluecmark& \bluecmark& \bluecmark \\
$\sigma_{\max}$  &    \nv   &     \nv    &   \nv    &  \redxmark &  \redxmark & \redxmark & \bluexmark & \bluexmark & \bluexmark & \bluexmark & \bluecmark & \redxmark & \bluecmark \\
$\sigma(\%)$   &    \nv   &    \nv     &    \nv   &   \nv & \redxmark &  \redxmark & \redxmark & \bluexmark & \redxmark & \redxmark & \bluecmark& \redxmark & \bluecmark \\
SMV           &     \nv  &    \nv     &  \nv     &   \nv & \nv & \redxmark & \bluexmark & \bluexmark & \bluexmark & \bluexmark & \bluecmark & \redxmark & \bluecmark \\
UEF          &    \nv   &    \nv     &    \nv   &  \nv  &   \nv& \nv & \bluexmark & \bluexmark & \bluexmark & \bluexmark & \bluecmark & \redxmark & \bluecmark \\
RSD          &   \nv    &    \nv     &     \nv  &  \nv  &   \nv&   \nv & \nv& \bluexmark &  \redxmark & \bluexmark & \bluecmark& \redxmark & \bluecmark \\
QPP-PRP     &   \nv    &    \nv     &     \nv  &  \nv  &   \nv&   \nv & \nv & \nv & \redxmark & \redxmark & \redcmark& \redxmark & \bluecmark \\
SCNQC       &   \nv    &    \nv     &     \nv  &  \nv  &   \nv&   \nv & \nv & \nv  & \nv & \bluexmark & \bluecmark & \redxmark& \bluecmark\\
QV-NQC      &   \nv    &    \nv     &     \nv  &  \nv  &   \nv&   \nv & \nv & \nv & \nv & \nv & \bluecmark& \redxmark & \bluecmark \\
DM        &   \nv    &    \nv     &     \nv  &  \nv  &   \nv&   \nv & \nv & \nv & \nv & \nv  &\nv & \redcmark & \bluexmark\\
NQAQPP      &   \nv    &    \nv     &     \nv  &  \nv  &   \nv&   \nv & \nv & \nv & \nv & \nv & \nv &\nv & \bluecmark \\
BERTQPP     &   \nv    &    \nv     &     \nv  &  \nv  &   \nv&   \nv & \nv & \nv & \nv &  \nv&\nv &\nv & \nv\\
\end{tabular}
\end{adjustbox}
\caption{\local}
\label{tab:ttest_loca}
\end{subtable}
\caption{
%
Statistical significance comparisons ($t$-test, $p<0.05$) between pairs of QPP models under the \stand~and \local~setups. For each cell defined by $\text{row}=\phi_i$ and $\text{col}=\phi_j$, a \rbox{\tsquare} indicates that $\phi_i$ outperforms $\phi_j$ (i.e., $\mathcal{P}(\phi_i) > \mathcal{P}(\phi_j)$), while a \bbox{\tsquare} indicates the opposite. A `\ding{51}' indicates statistical significance. \textbf{Observations}: \textbf{(a)} a larger number of models significantly outperform others; \textbf{(b)} the two best-performing models (BERT-QPP and DM) significantly outperform all competitors, whereas the weakest model (WIG) is significantly outperformed by every other model.
%
%
}
\label{tab:ttests_ap}
\end{table*}